\documentclass[
reprint,
amsmath,amssymb,
aps,
nofootinbib,
pre,
twocolumn
]{revtex4-1}

\usepackage{mathbbol}
\usepackage{graphicx}
\usepackage{dcolumn}
\usepackage{bm}

\newcommand{\la}[1]{\label{#1}}
\newcommand{\eq}[1]{(\ref{#1})}
\newcommand{\commented}[1]{}

\newcommand{\beq}{\begin{equation}}
\newcommand{\eeq}{\end{equation}}
\newcommand{\beqq}{\begin{equation*}}
\newcommand{\eeqq}{\end{equation*}}
\newcommand\beqa{\begin{eqnarray}}
\newcommand\eeqa{\end{eqnarray}}
\newcommand\beqaa{\begin{eqnarray*}}
\newcommand\eeqaa{\end{eqnarray*}}
\newcommand\bea{\begin{array}}
\newcommand\eea{\end{array}}

\def\tr{{\rm tr}\;}

\newcommand{\hs}{\frac{\sqrt{3}}{2}}

\newcommand{\<}{{\langle}}
\renewcommand{\>}{{\rangle}}

\def\({\left(}
\def\){\right)}
\def\[{\left[}
\def\]{\right]}

\def\<{\langle}
\def\>{\rangle}

\def\bB{{\bf B}}
\def\bC{{\bf C}}

\def\mC{{\mathbb C}}
\def\mB{{\mathbb B}}

\def\s*{\ *_{\!\!\!\!\!\!\!\!\!\,_{\,_\text{\scriptsize{sym}}}}}
\def\hs*{\ \hat{*}_{\!\!\!\!\!\!\!\!\!\,_{\,_\text{\scriptsize{sym}}}}}

\begin{document}

\begin{flushright}
 
\end{flushright}
\vspace{-5truecm}
\title{Dual Separated Variables and Scalar Products}

\author{Nikolay Gromov}
\email{nikolay.gromov@kcl.ac.uk}
\affiliation{
Mathematics Department, King's College London,
The Strand, London WC2R 2LS, UK
}
\affiliation{St.Petersburg INP, Gatchina, 188 300, St.Petersburg,
Russia}
\vspace*{-250px}

\author{Fedor Levkovich-Maslyuk}
\email{fedor.levkovich@gmail.com}
\affiliation{
Departement de Physique, Ecole Normale Superieure / PSL Research University, CNRS, 24 rue Lhomond, 75005 Paris, France
}
\affiliation{Institute for Information Transmission Problems, Moscow 127994, Russia
}
\vspace*{-250px}

\author{Paul Ryan}
\email{pryan@maths.tcd.ie}
\affiliation{School of Mathematics \& Hamilton Mathematics Institute, Trinity College Dublin, College Green, Dublin 2, Ireland
}
\vspace*{-250px}

\author{Dmytro Volin}
\email{dmytro.volin@physics.uu.se}
\affiliation{Nordita, KTH Royal Institute of Technology and Stockholm University, Roslagstullsbacken 23, SE-106 91 Stockholm, Sweden
}
\affiliation{Department of Physics and Astronomy,
Uppsala University, Box 516, SE-751 20 Uppsala, Sweden
}
\vspace*{-250px}

\begin{abstract}
Separation of variables (SoV) is an extremely efficient and elegant technique for analysing physical systems but its application to integrable spin chains was limited until recently to the simplest $\mathfrak{su}(2)$ cases. In this paper we continue developing the SoV program for higher-rank spin chains and demonstrate how to derive the measure for the $\mathfrak{su}(3)$ case. Our results are a natural consequence of factorisability of the wave function and functional orthogonality relations following from the interplay between Baxter equations for Q-functions and their dual.

\end{abstract}

\pacs{Valid PACS appear here}
\maketitle

\section{\label{intro}Introduction}
The key physical information contained in a quantum system is encoded into matrix elements of operators between Hamiltonian eigenstates, but computing them is not a simple task. 
To begin with one should carefully choose a coordinate system. Famously, in the case of the hydrogen atom the problem greatly simplifies in spherical coordinates -- the wave function splits into six
 independent one-dimensional factors which allows one to perform many computations analytically.

A possible price to pay for such a simple factorised form of the wave function could come from a complicated integration measure in the scalar product. In the case of the hydrogen atom it is simply $r^2 \sin\theta$, but the problem can become rather challenging in general. In this paper we address it for integrable spin chains.

Like the hydrogen atom, many integrable models are believed to admit a separation of variables (SoV) basis, where the wave function becomes a product of simple factors. A particularly important model is the $\mathfrak{su}(2)$ Heisenberg spin chain which is a model of interacting particles on a one-dimensional chain of sites. In its simplest realisation, its Hamiltonian is given by $H=-J\sum_\alpha \vec\sigma_\alpha\vec \sigma_{\alpha+1}$,
where $\vec \sigma_\alpha$ are the Pauli matrices acting on the site $\alpha$. This model is known to be integrable and the separation of variables was  worked out by Sklyanin in \cite{Sklyanin:1984sb,SklyaninFBA}.

The integrable structures of a given model greatly depend on the underlying symmetry of the system. In recent years, there has been a great interest in studying integrable systems with more
general $\mathfrak{su}(N)$-symmetries and super-symmetries coming from
the AdS/CFT correspondence and integrability in string theory.
In particular, the Fishnet model \cite{Zamolodchikov:1980mb,Gurdogan:2015csr}  is essentially an $\mathfrak{su}(4)$
rational spin chain and ${\cal N}=4$ SYM is tightly related to the $\mathfrak{psu}(2,2|4)$ integrable spin chain. The challenge of computing correlation functions in these models provides important motivation for developing the SoV approach in more complicated systems. In this paper we consider the case of $\mathfrak{su}(3)$ spin chains, and many of our techniques can be carried over to $\mathfrak{su}(N)$. 

The general $\mathfrak{su}(N)$ Heisenberg spin chain of length $L$ is defined by means of the Hamiltonian $H=\sum_{i=1}^L\mathcal{H}_{i,i+1}$ where the nearest-neighbour Hamiltonian density $\mathcal{H}_{i,i+1}$ is simply the permutation operator $P_{i,i+1}$ permuting the $\mathfrak{su}(n)$ ``spins" at each site, and one usually assumes periodic boundary conditions $\mathcal{H}_{L,L+1}=\mathcal{H}_{L1}$. A more algebraic formulation of the model is in terms of the $R$-operator $R^{a q}_{bp}(u)=(u-\tfrac i2)\delta^a_b\delta^q_p+ i\delta^a_p\delta^q_b$ which allows one to uncover the integrable structures of the model and build a tower of conserved charges in a constructive way, as well as introduce some extra parameters. To this end we further build the monodromy matrix
\beq
\hat T^{a}{}_{b}=R^{a\;\, q_1}_{c_1 p_1}(u-\theta_1)\;
R^{c_1 q_2}_{c_2 p_2}(u-\theta_2)\dots
R^{c_{L-1} q_L}_{b\;\;\;\;\;\; p_L}(u-\theta_L)z_b\,,
\nonumber
\eeq
where we assume summation over all repeated indices except $b$ and have also introduced the parameters $\theta_\alpha$, known as the inhomogeneities, and the twist parameters $z_j$, and without loss of generality we can assume $z_1z_2\dots z_N=1$. The inhomogeneities $\theta_\alpha$ break the local structure of the Hamiltonian and the twist modifies the boundary conditions. Their benefit comes from the fact that they remove certain degeneracies present in the model allowing for more straightforward calculations. 

The monodromy matrix is a collection of $N^2$
operators $\hat T^a{}_{b}(u)$ each acting on the physical Hilbert space $(\mathbb{C}^N)^{\otimes L}$. The trace of the monodromy matrix, $\hat t(u)=\tr \hat T(u)$, known as the transfer matrix, forms a family of mutually commuting operators $[\hat t(u),\hat t(v)]=0$, and so these operators are obtained as the coefficients of the spectral parameter $u$ in the expansion of the matrix $\hat t(u)$. Unfortunately, for $\mathfrak{su}(N)$ this set of conserved charges is not maximal. To obtain more conserved charges one should take the trace of the monodromy matrix in other representations of $\mathfrak{su}(N)$, in particular one can generate all conserved charges by restricting to antisymmetric representations of which there is only a finite number $N-1$, and we denote the corresponding transfer matrices by $\hat t_a(u)$, $a=1,\dots,N-1$. We restrict ourselves to $\mathfrak{su}(3)$ in this paper which is general enough to illustrate our construction while allowing for a relative clarity. In this case we have the fundamental representation and the twice antisymmetric one, with the corresponding transfer matrices being
\beq
\hat t_1(u)=\tr \hat T(u)\;\;
,\;\;\hat t_2(u)=\tr \hat U(u-i)\;,
\eeq
where
$
\hat U_{a}{}^{b}(u)=\frac{1}{2}
\epsilon_{a a_1 a_2}
\epsilon^{b b_1 b_2}
\hat T^{a_1}{}_{b_1}(u)\hat T^{a_2}{}_{b_2}(u+i)$.
We see that $\hat t_2(u)$ is a polynomial in $u$ of degree $2L$. However, $\hat U$ contains a trivial factor $Q_{\theta}(u-\tfrac{3i}{2})$
where $Q_\theta(u)\equiv\prod_{\alpha=1}^L(u-\theta_\alpha)$ and so $\hat t_2$ generates only $L$ new commuting operators. In the following we use \beq
\label{taudef}
\hat\tau_1(u)=\hat t_1(u)\;\;,\;\;
\hat\tau_2(u)=\frac{\hat t_2(u)}{Q_\theta(u-\tfrac{3i}2)}\,.
\eeq
The same quantities without hats will denote the eigenvalues. In the next section we review how they can be computed using integrability. Then in sections III and IV we discuss the construction of wavefunctions in the basis of separated variables and present our main results.

\section{Baxter Q-functions}
The integrability of the model promises a number of simplifications.
In particular, its spectrum can be computed relatively easily.
The modern way of approaching the spectral problem is via Q-functions \cite{Baxter,Bazhanov:1996dr,Krichever:1996qd} (also known as Baxter polynomials). We summarise the key results of this approach in this section, see e.g. \cite{Kazakov:2015efa,Gromov:2010kf} for a comprehensive review. In the next sections we will see that the Q-functions are also very convenient building blocks for the spin chain wave functions. 

The basic Q-functions are the {\it twisted} polynomials $q^j(u),\  j=1,2,3$, {\it i.e.}
polynomial functions up to an exponential prefactor, of the form $q^j(u)=z_j^{iu}(u^{M_j}+\ldots)$, where $M_j$ obey $M_1+M_2+M_3=L$.
An alternative to the widely used nested Bethe equations and in many ways better method of finding the spectrum of the system is to impose the {\it quantization condition} 
\beq\la{quant}
\epsilon_{ijk}q^i(u-i)q^j(u)
q^k(u+i)\propto Q_\theta(u)\;.
\eeq
The condition \eq{quant} gives $L$ equations on the total $L$ roots of $q^i(u)$, selecting the physical solutions. 
One advantage w.r.t. the conventional Bethe ansatz is that it allows one to count solutions more easily. For example, when all $|\theta_i-\theta_j|$ are large, \eqref{quant} reduces to $q^1q^2q^3=Q_{\theta}$ which has $3^L$ solutions, {\it i.e.} equal to the dimension of the Hilbert space. As the dependence on the parameters $\theta_\alpha$ should be continuous, except probably for some special points, this 
leads to the completeness of the equation \eqref{quant}.
For more detailed and mathematically rigorous discussion see \cite{MTV,VolinLeurentChernyak}. In these works it is shown that the algebraic number of solutions to \eqref{quant} is equal to $3^L$ for any numerical values of inhomogeneities $\theta_{\alpha}$.

To relate the quantisation condition \eqref{quant} to conventional Bethe equations we also need dual functions $q_i$ introduced as
\beq\la{qidef}
q_i(u)\propto \epsilon_{ijk}q^j(u+\tfrac{i}{2})
q^k(u-\tfrac{i}{2})\;.
\eeq
The normalization coefficient in \eq{qidef} is such that $q_j(u)=z_j^{-i u}(u^{L-M_j}+\dots)$.

One can take any of the six choices of $q^j$ and $q_i$ such that $i\neq j$. This maps to six possible nested Bethe equations for twisted spin chains that differ by the choice of a ``ferromagnetic" vacuum: zeros of $q_i$ are the so-called momentum-carrying roots and zeros of $q^j$ are the so-called auxiliary Bethe roots.

Transfer matrices ${\tau}_a$ defined in \eq{taudef} can be reconstructed from the Q-functions using simple contractions
\begin{align}\la{tq}
\tau_1 &\propto q^j(u+\tfrac {3i}{2})\, q_j(u-i)\,,
&\!\!
\tau_2 &\propto q^j(u-\tfrac {3i}{2})q_j(u+i)\,.
\end{align}
The last formula suggests that $\tau_i$ are Hermitian conjugates of one another which is indeed the case if the twists $z_j$ are pure phases and the inhomogeneities $\theta_{\alpha}$ are real. Finally, we shall later need the following special values of $\tau_a(u)$ following from \eq{tq}:
\beq\la{tspecial}
\tau_2(\theta_\alpha-\tfrac{i}{2})=Q_\theta(\theta_\alpha-i)\frac{q_1(\theta_\alpha+\tfrac{i}{2})}{q_1(\theta_\alpha-\tfrac{i}{2})}\;,
\eeq
and
\beqa\la{tspecial2}
\!\!\!\!\!\!\!\!\frac{\tau_1(\theta_\alpha-\tfrac{i}{2})}{{Q_\theta(\theta_\alpha-i)}}
\!\!=\!\! \frac{
q^2(\theta_\alpha\!\!-i)
q^3(\theta_\alpha\!\!+i)
\!-\!
q^3(\theta_\alpha\!\!-i)
q^2(\theta_\alpha\!\!+i)
}{
q^2(\theta_\alpha\!\!-i)
q^3(\theta_\alpha)
-
q^3(\theta_\alpha\!\!-i)
q^2(\theta_\alpha)
}\;.
\eeqa

\section{Separation of Variables}
\paragraph{SoV basis.}
Let us summarize some of the known results concerning the implementation of the SoV approach for quantum spin chains.
Motivated by the SoV construction in the classical limit \cite{Sklyanin:1992eu}, Sklyanin realised  in \cite{Sklyanin:1992sm} that the operator
\beq\la{Bdef}
\hat{\mB}(u)=\hat T^{2}{}_{3}(u)\hat U_{3}{}^{1}(u-i)
-\hat T^{1}{}_{3}(u)\hat U_{3}{}^{2}(u-i)
\eeq
should play an important role in {\it{quantum}} separation of variables for the model. However, the precise understanding of how the SoV procedure should work was only recently obtained in \cite{Gromov:2016itr}, where several important observations were made: 
Firstly, Sklyanin's construction remains intact under the replacement $\hat T\to \hat T^g\equiv  g^{-1}\hat T g$, where $g$ is some constant $\mathsf{SL}(3)$ matrix. This replacement makes $\hat{\mB}(u)$ diagonalisable for generic enough $g$ \footnote{For definiteness one can take $g_{pq}=1$ except for $g_{21}=g_{32}=0$. A similar observation for a  model with $\mathfrak{su}(2)$ symmetry was also made in \cite{Sklyanin:1989cg}.} 
and so its spectrum and eigenvalues become interesting quantities to consider. Secondly, the spectrum of $\hat {\mB}^g(u)$ is non-degenerate and has the following remarkably regular structure. Namely, for $\hat\mB^g=\Lambda\, {\hat{\bf B}^g}$, where $\Lambda = \Lambda_0 Q_\theta(u-3i/2)$ is a trivial scalar factor that does not depend on the state, the eigenvalues of $\hat {\bf B}^g$ are given by
\beq\la{BGei}
{\bf B}^g(u)=\prod_{\alpha=1}^L\prod_{a=1}^2(u-\theta_\alpha-\tfrac{i}{2}+im_{\alpha,a})\,,
\eeq
where the integers $m_{\alpha,a}$ satisfy  $0\leq m_{\alpha,1}\leq m_{\alpha,2}\leq 1$. 

The operators $\hat\mB(u)$ commute with each other for different values of $u$ \cite{Sklyanin:1992sm}.
The same holds true for $\hat\mB^g(u)$ and thus eigenstates of $\hat\mB^g(u)$ do not depend on $u$. We denote its left eigenstates as $\langle x|$, labelling them by the values of $m_{\alpha,a}$. One can then unambiguously define $2L$ commuting operators $\hat {\bf X}_{\alpha,a}$ such that $\hat{\bf B}^g(u)=\prod_{\alpha}(u-\hat{\bf X}_{\alpha,1})(u-\hat {\bf X}_{\alpha,2})$
with eigenvalues being ${\bf X}_{\alpha,a}=\theta_\alpha-\frac{i}{2}+i\,m_{\alpha,a}$.

Finally, it was observed in \cite{Gromov:2016itr}
that the eigenstates of transfer matrices
can be constructed using the operator $\hat{\bf B}^g(u)$ as follows \beq\la{statecr}
|\Psi_n\> = 
\hat {\bf B}^g(u_1)\dots
\hat {\bf B}^g(u_{M_1})
|\Omega\>\;,
\eeq
where $u_i$ are the roots of the twisted polynomial $q_1(u)$ and $|\Omega\>=\delta^{p_1}_1\delta^{p_2}_1\dots \delta^{p_L}_1$
is a ``ferromagnetic vacuum" of the model. 

By combining \eq{statecr} with the definition of $\hat {\bf X}_{\alpha,a}$ we get a factorized representation of the wave function~\cite{Gromov:2016itr}
\beq\la{Psifac}
\Psi_n(x)\equiv \langle x|\Psi_n\rangle = \prod_{\alpha=1}^L q_1({\bf X}_{\alpha,1})q_1({\bf X}_{\alpha,2})\;,
\eeq 
and so $\langle x|$ form an SoV basis. In \eq{Psifac} we impose the following normalization for $\<x|$ s.t.
$
\langle x|\Omega\rangle = \prod_{a,\alpha}z_1^{-i\bf X_{\alpha,a}}
$.

While some of the observations of \cite{Gromov:2016itr} were conjectured based on numerical evidence or for short spin chains or small number of magnons, they received a complete analytical proof in \cite{Liashyk:2018qfc,Ryan:2018fyo}. In particular, it became clear that the spectrum of $\hat\mB^g$ given in \eq{BGei} originates from the structure of the Gelfand-Tsetlin algebra \cite{Ryan:2018fyo}. It would be interesting to examine if such a structure is also present in the separated variables considered in \cite{smirnov}.

An important observation can be made about the action of the transfer matrices at special values: $\frac{\hat \tau_2(\theta_\alpha-\tfrac i2)}{Q_\theta(\theta_\alpha-i)}$. Due to the relation
\eq{tspecial} it is clear that acting on the state $\<x|$
they would replace one factor of $q_1(\theta_\alpha-\tfrac{i}{2})$
by $q_1(\theta_\alpha+\tfrac{i}{2})$ in the r.h.s. of \eq{Psifac} and thus they play the role of the creation operators for the basis $\<x|$~\cite{GSFup}. More precisely
\beq \label{Bstate}
\<x| = \<0|\prod_{\alpha=1}^L \(\frac{\hat \tau_2\left(\theta_\alpha-\tfrac{i}{2}\right)}{
Q_\theta(\theta_\alpha-i)
}\)^{{m_{\alpha,1}}+{m_{\alpha,2}}}\,,
\eeq 
where $\<0|$ is the eigenstate of $\hat\bB^g$ with all $m_{\alpha,a}=0$.
This observation demonstrates the equivalence with a more recent approach of~\cite{Maillet:2018bim}, where an analog of \eqref{Bstate} was taken as the starting point, and it generalises beyond fundamental representation \cite{Ryan:2018fyo}.
In the approach of~\cite{Maillet:2018bim} one can avoid discussing completeness of the quantization conditions, such as Bethe equations. While for original Bethe equations completeness is a notorious obstacle, using the elegant condition \eqref{quant} instead removes this difficulty.

\paragraph{Dual SoV basis.} 
Now that we have reviewed the results previously established in the literature we come to our new findings. Since our main objective in this work is the computation of scalar products $\langle \Psi|\Psi\rangle$ in the SoV framework, the next ingredient we need is an SoV representation of the left eigenstates $\langle \Psi|$. Let us describe its realization, which is one of our main new results. Naively, one could try to use the right eigenstates $|x\rangle$ of $\hat \bB^g$ but unfortunately this does not work as the resulting wave function $\langle \Psi|x\rangle$ does not factorise. To circumvent this problem it proves fruitful to perform dualisation of the monodromy matrix instead. This is done by using the so-called antipode map which sends the monodromy matrix considered as a $3\times 3$ matrix with non-commutative entries $\hat T^a{}_b$ to its inverse. To explicitly compute the inverse we notice that $\hat U^T$ looks like the adjunct matrix for $\hat T$ and, indeed, it satisfies a quantum analog of the Cramer's formula $U_{b}{}^{a}(u-i)T^{b}{}_{c}(u+i)=Q_{\theta}(u+\tfrac {3i}{2})Q_{\theta}(u-\tfrac i2)Q_{\theta}(u-\tfrac {3i}2)\delta_{ac}.$ Employing it we compute how $\hat\mB(u)$ transforms under the antipode map and obtain, with convenient adjustment of normalisation and shift of $u$, a new operator 
\beq\la{Cdef}
\hat \mC(u)=
\hat T^2{}_{3}(u-\tfrac i2)\hat U_{3}{}^{1}(u-\tfrac i2)
-\hat T^1{}_{3}(u-\tfrac i2)\hat U_3{}^{2}(u-\tfrac i2)
\;
\eeq
which is one of the main results of this paper\footnote{Curiously, a similar operator also denoted $C(u)$ appears at an intermediate step of a technical calculation in \cite{Sklyanin:1992sm}. However, none of its crucial properties that we describe here were discussed there.
}. Remarkably, the only difference between $\hat \mC(u)$ and $\hat \mB(u)$ is in the shifts of the spectral parameter, meaning that there is no difference in the classical limit where shifts are ignored.

We found that essentially the same facts are true for $\hat\mC(u)$ as for $\hat\mB(u)$.  We again perform the replacement trick $\hat\mC(u)\to\hat\mC^g(u)$ and introduces $\bC^g$ by removing the trivial non-dynamical factor, $\bC^g(u)\propto\hat\mC^g(u)/Q_\theta(u-i)$. Due to the commutativity $[\hat\bC^g(u),\hat\bC^g(v)]=0$, $\hat\bC^g(u)$ has $u$-independent eigenvectors dubbed $|y\>$. Furthermore, this right basis $|y\>$ does indeed factorise the left eigenfunctions $\langle \Psi|$ of the transfer matrices.

The spectrum of $\hat\bC^g(u)$ is of the form
\beq
\bC^g(u)=\prod_{\alpha=1}^L\left(u-\theta_\alpha-i n_{\alpha,2}\right)\left(u-\theta_\alpha+i-i n_{\alpha,1}\right)\,,
\eeq
where $0\leq n_{\alpha,1}\leq n_{\alpha,2}\leq 1$.

The eigenstates $|y\>$ can also be built in the spirit of \eq{Bstate} but in a slightly modified form, similar to the construction of \cite{Ryan:2018fyo} for a spin chain in the anti-fundamental instead of fundamental representation. 
Indeed, we found that the results of~\cite{Ryan:2018fyo}
apply but for the right eigenstates
\beq \label{Cstate}
|y\> = \prod_{\alpha=1}^L \frac{\hat \tau_1\left(\theta_\alpha-\tfrac{i}{2}\right)^{{n_{\alpha,2}}-{n_{\alpha,1}}}\hat \tau_2\left(\theta_\alpha-\tfrac{i}{2}\right)^{n_{\alpha,1}}}{[Q_\theta(\theta_\alpha-i)]^{n_{\alpha,2}}}|0\rangle\;,
\eeq 
where $|0\rangle$ is the eigenvector of $\hat\bC^g$ with $n_{\alpha,a}=0$. Since the proof is technical and identical to that of \cite{Ryan:2018fyo} we do not reproduce it here.

We then introduce another set of separated variables $\hat {\bf Y}_{\alpha,a}$ by specifying their eigenvalues on the above states, namely by
\beq
\hat \bC^g(u)=\prod_{\alpha}(u-\hat {\bf Y}_{\alpha,1})
(u-\hat {\bf Y}_{\alpha,2})\,,
\eeq
where ${\bf Y}_{\alpha,1}=\theta_\alpha-i+in_{\alpha,1}$, ${\bf Y}_{\alpha,2}=\theta_\alpha+in_{\alpha,2}$.
With these variables at hand, we factorise the transfer matrix eigenstates $\<\Psi|$ exactly as it was done for
$|\Psi\rangle$ in \cite{Ryan:2018fyo} for the anti-fundamental representation. By computing the overlap $\langle \Psi|y\rangle$ and using \eqref{Cstate} we obtain
\beq
\langle \Psi|y\rangle=
\prod_{\alpha=1}^L  \frac{\tau_1\left(\theta_\alpha-\tfrac{i}{2}\right)^{{n_{\alpha,2}}-{n_{\alpha,1}}} \tau_2\left(\theta_\alpha-\tfrac{i}{2}\right)^{n_{\alpha,1}}}{[Q_\theta(\theta_\alpha-i)]^{n_{\alpha,2}}}
\<\Psi|0\rangle\,.
\eeq
Next, normalizing the states $\<\Psi|$ so that 
\beq
\langle \Psi|0\rangle=
\prod_{\alpha=1}^L\[q^2(\theta_\alpha-i)q^3(\theta_\alpha)
-
q^3(\theta_\alpha-i)q^2(\theta_\alpha)\]
\eeq
and using \eq{tspecial} and \eq{tspecial2}, we conclude
\beq\la{SoVright}
\langle \Psi|y\rangle=
\prod_{\alpha=1}^L\[q^2({\bf Y}_{\alpha,1})q^3({\bf Y}_{\alpha,2})
-
q^3({\bf Y}_{\alpha,1})q^2({\bf Y}_{\alpha,2})\]\;.
\eeq
\paragraph{SoV-charge operator.}
Since the operators $\bB^g(u)$ and $\bC^g(u)$ only differ by
shifts in their definitions \eq{Bdef} and \eq{Cdef}, they become related at large $u$. In particular, their first two terms of the large-$u$ expansion are exactly the same. While the leading term is proportional to the identity matrix, the subleading coefficient
defines the {\it SoV-charge operator} $\hat{\bf S}$
\beq
\hat\bC^g(u)\simeq \hat\bB^g(u)\simeq u^{2L}+u^{2L-1}\[\sum_{\alpha=1}^L(2\theta_\alpha-i)+i \hat{\bf S}\]\;.
\eeq
$\hat{\bf S}$ commutes with both $\hat\bB^g(u)$ and
$\hat\bC^g(u)$ by construction, and it counts the number of ``excitations" in the SoV states:
\beq
{\bf S}=\sum_{\alpha,a}n_{\alpha,a}=\sum_{\alpha,a}m_{\alpha,a}
\eeq
which will be a useful observation later.

\paragraph{Scalar product in the SoV basis.} 
Our goal is to express the scalar product in SoV variables in a closed form.
For any two bases $|y\>$ and $\<x|$ one can write
\beq\la{scalar}
\<\Phi|\Phi'\> = 
\sum_{x}\sum_y{\cal M}_{x,y}\<\Phi|y\>\<x|\Phi'\>\,,
\eeq
where the {\it measure} ${\cal M}_{x,y}$ is the inverse transposed matrix of the overlaps $\<x|y\>$. Without making any calculation, we can make an important observation about the matrix  $\<x|y\>$ -- the existence of the SoV-charge operator $\hat {\bf S}$ implies that only the matrix elements with the same excitation numbers $\sum n_{\alpha,a}=\sum m_{\alpha,a}$ can be non-zero. In particular, the ground state $\<0|$ should be also an eigenstate of $\bC(u)$, and, as the spectrum of $\bC(u)$ is non-degenerate, this means that $\<0|y\>\propto\delta_{0,y}$
and similarly $\<x|0\>\propto\delta_{x,0}$, which also implies that ${\cal M}_{x,0}\propto \delta_{x,0}$ and ${\cal M}_{0,y}\propto \delta_{0,y}$. 

When $\hat\tau_1$ and $\hat\tau_2$ (as operators) can be diagonalised their joint spectrum must be non-degenerate \cite{MTV} and then their joint left and right eigenstates are orthogonal $\langle \Psi_A|\Psi_B\rangle={\cal N}_A^2\delta_{AB}$. Using the SoV basis, we then have
\beqa\la{ortN}
&&{\cal N}^2_A\delta_{AB}=
\sum_{x,y}{\cal M}_{x,y}
\prod_{\alpha=1}^L q^A_1({\bf X}_{\alpha,1})q^A_1({\bf X}_{\alpha,2})\\
\nonumber&&\times\prod_{\alpha=1}^L\[q_B^2({\bf Y}_{\alpha,1})q_B^3({\bf Y}_{\alpha,2})
-
q_B^3({\bf Y}_{\alpha,1})q_B^2({\bf Y}_{\alpha,2})\]\;,
\eeqa
where $q^A_j$ and $q_B^j$ are the Q-functions corresponding to the
eigenstate $\Psi_A$ and $\Psi_B$.

\section{Functional Orthogonality Relation}
Now we shall consider the orthogonality question and reproduce \eqref{ortN} following the method of \cite{Cavaglia:2019pow,Cavaglia:2018lxi}. The starting point is the two Baxter TQ-relations.
With the help of the finite difference operator
\beq
{O}=\frac 1{Q_{\theta}^-}D^{-3}-\frac{\tau_2}{Q_{\theta}^+Q_{\theta}^-}D^{-1}+\frac{\tau_1}{Q_{\theta}^+Q_{\theta}^-}D-\frac 1{Q_{\theta}^+}D^{+3}\,,
\eeq
where $D\equiv e^{i/2\partial_u}$, both Baxter relations are written in a unified way
$$
\overrightarrow{O}q^i=0\;, \quad{\rm and}\quad q_i\overleftarrow{ O}=0\,,
$$
where arrows indicate the direction in which the shift operator acts: $\overrightarrow{D}f=f(u+i/2)$ and $g\overleftarrow{D}=g(u-i/2)$.

The orthogonality conditions shall be now built using the following simple fact $\oint \mu(u) (f \overrightarrow{O} g)du=\oint \mu(u) (f\overleftarrow{O} g)$, where the measure $\mu(u)$ is an $i$-periodic analytic function, $f$ and $g$ are analytic and the contour is a large enough circle which is easily demonstrated by shifting the contour of integration.
In particular we have
\beq\la{bp}
0=\oint   q^A_{i}\; \overleftarrow O_A\;q_B^j  e^{2\pi u \beta}du=
\oint   q^A_{i}\; \overrightarrow O_A \;q_B^j e^{2\pi u \beta} du
\;,
\eeq
where $\beta\in{\mathbb Z}$ and the indices $A$ and $B$ indicate 
the eigenstates of the transfer matrix. Note that the finite difference operator $O_B$ itself depends on these states through the eigenvalues $\tau_a$ of the transfer matrices.
The integrand has $2L$ poles at $\theta_i\pm \frac{i}{2}$.
These poles are cancelled by the trigonometric polynomial
$\prod_i^L(e^{2\pi u \beta}+e^{2\pi \theta_i \beta})$ (defined in such a way that it has zeros at these points and is an $i$-periodic function of $u$), meaning
that there are only $L$ linearly independent exponents one can insert and thus one can restrict
$\beta=1,\dots,L$. From \eq{bp} we obtain
$
\oint   q^A_i\[\overrightarrow O_B-\overrightarrow O_A\]  q_B^{j}e^{2\pi u \beta} du = 0
$, or
\beq
\oint   q^A_i
\frac{\Delta\tau_2 q_B^{j}(u-\tfrac i2)
+
\Delta\tau_1 q_B^{j}(u+\tfrac i2)}{Q_{\theta}^+Q_{\theta}^-}e^{2\pi u \beta} du = 0\,,
\eeq
where $\Delta\tau_a=(-1)^a(\tau_a^A-\tau_a^B)=\sum_{\alpha} \Delta I_{a,\alpha} u^{\alpha-1}$. We take $i=1$ and $j=2,3$, which gives 
\beq
\sum_{a,\alpha}\Delta I_{a,\alpha}
\oint   
\frac{q^A_1(u) u^{\alpha-1} q_B^{1+b}(u+\tfrac {3i}2-ia)}{
Q_{\theta}(u+\tfrac i2)Q_{\theta}(u-\tfrac i2)
}e^{2\pi u \beta} du=0\;,
\eeq
where $b=1,2$ and $\beta=1,\dots, L$, giving in total $2L$ equations.
Consider it as a linear system on $\Delta I_{a,\alpha}$. To have a non-trivial solution it must be degenerate, meaning that if $A$ and $B$ are different states we have ${\rm det}\; M=0$,  where
\beq\la{qqo}
M_{(a,\alpha),(b,\beta)}=
\oint   
\frac{q^A_1(u) u^{\alpha-1} q_B^{1+b}(u+\tfrac {3i}2-ia)}{
Q_{\theta}(u+\tfrac i2)Q_{\theta}(u-\tfrac i2)
}e^{2\pi u \beta} du\;.
\eeq
The equation $\det\;M=0$ (for $A\neq B$) is the functional orthogonality relation. To relate it to our operatorial SoV construction we compute the integral by residues. If one first performs a simple linear transformation $e^{2\pi u\beta} \to \prod_{\gamma\neq \beta} \left(e^{2\pi u}+e^{2\pi\theta_\gamma}\right)$, which changes $M\to \tilde M$ but does not affect the determinant value, the new $i$-periodic factor would cancel all the poles except the ones at $u=\theta_{\beta}\pm \tfrac{i}{2}$ and the result of integration is $\tilde M_{(a,\alpha),(b,\beta)}$ equal to
\beqa
\nonumber&&+q_1^A(\theta_\beta+i/2)\;
\frac{(\theta_\beta+i/2)^{\alpha-1}}{\prod_{\gamma} {(\theta_\beta-\theta_\gamma+i)}}\;  q_B^{1+b} \(\theta_\beta-ai+2i\)\\
\nonumber&&+q_1^A(\theta_\beta-i/2)\;
\frac{(\theta_\beta-i/2)^{\alpha-1}}{\prod_{\gamma} {(\theta_\beta-\theta_\gamma-i)}}\;  q_B^{1+b} \(\theta_\beta-ai+i\)\;.
\eeqa
Let us see that $\det\tilde M(A,B)$ has exactly the form of the r.h.s. of \eq{ortN}!
Indeed, we are guaranteed to get a sum of terms each containing a product of $2L$ factors $q_1^A(\theta_\beta\pm i/2)$. Now, if we fix
some combination of $2L$ $\pm$ signs, we are left with a determinant containing $q_B^{1+b}(\theta_\beta \pm i)$ and $q_B^{1+b}(\theta_\beta)$ with dependence on $b$ contained only in the index of the Q-function, meaning that 
the final expression will be anti-symmetrized in $b$ for each given $\beta$, but the only antisymmetric in $b$ combinations of $q_B$'s are the factors of the type $q_B^2({\bf Y}_{\beta,1})q_B^3({\bf Y}_{\beta,2})
-
q_B^3({\bf Y}_{\beta,1})q_B^2({\bf Y}_{\beta,2})$. The remaining coefficients are some combinations of $\theta$'s.

Now we show that
\beq\la{final}
\det\tilde M(A,B)=\langle \Psi_A|\Psi_B\rangle\,,
\eeq
up to an overall rescaling of $\tilde M$.

To this end consider the equation \eq{ortN} for $A\neq B$
as a set of $3^L \times (3^L-1)$ linear equations on 
$3^L\times 3^L$ quantities
${\cal M}_{x,y}$.
Furthermore, we can fix $3^L$ variables ${\cal M}_{x,0}=c \delta_{x,0}$, making the number of unknowns and equations to be the same. This means that, assuming independence of the   equations, there should be a unique solution for ${\cal M}_{x,y}$ up to an overall rescaling, parametrized by the constant $c$. Coefficients of the expansion of $\det\;\tilde M$ over the SoV wave functions \eqref{Psifac} and \eqref{SoVright} construct this solution for us. Finally, to fix the overall constant $c$ we can take \eq{ortN} for $\Psi^A=\Psi^B=\Omega$.
Using that the l.h.s. for this state is $1$ and all Q-functions are trivial, one can find the constant $c$ too.

We conclude that by using the orthogonality relations following from the Baxter TQ-equations we can completely fix the measure and thus obtain the scalar product in separated variables \eqref{final}\footnote{The result relies on the assumption of independence of the equations discussed in the previous paragraph. As a check, we explicitly verified it analytically for spin chains of small length.}. Note that \eqref{final} would hold even when both states are ``off-shell" that we define as states \eqref{Psifac} and \eqref{SoVright} but for $q_a$ and $q^a$ not satisfying the quantisation condition \eqref{quant}.

\section{Conclusions}
In this paper we constructed SoV bases for both bra and ket states, with a relatively simple overlap, providing a measure for the scalar product. We also showed how to find the SoV measure based on the method of \cite{Cavaglia:2019pow}, which bypasses an explicit operatorial computation and allows us to extract the result from a simple  determinant. 
In a similar way one can compute matrix elements of a large class of operators such as $\mB(u),\;\mC(u)$ and $\hat t_a(u)$, which are expected to generate the full algebra of observables.
Further generalisations of our results will be reported in \cite{GLRV} (in particular, the general $\mathfrak{su}(N)$ case can be addressed using the same techniques).
Finally, it would be interesting to understand the relation between \eq{final} and Gaudin norms~\cite{Hutsalyuk:2017dit} and recent results involving Gaudin matrices~\cite{Pozsgay:2019xak}.

{\it Acknowledgements:} 
We are grateful to  A.~Cavaglia, G.~Ferrando, M.~de~Leeuw, S.~Leurent, J.-M.~Maillet, G.~Niccoli, D.~Serban and F.~Smirnov for discussions. 
N.G. thanks Trinity College Dublin for hospitality during a part of this project.
F.L.-M. thanks Nordita Stockholm for hospitality during a part of this project.
This work is supported by the STFC grant (ST/P000258/1) and by Agence Nationale de la Recherche LabEx grant ENS-ICFP ANR-10-LABX-0010/ANR-10-IDEX-0001-02 PSL. The work of P.R. is supported in part by a Nordita Visiting PhD Fellowship and by SFI and the Royal Society grant UF160578. The work of D.V. is supported by the Knut and Alice Wallenberg Foundation under grant Dnr KAW 2015.0083.

\end{document}